\documentclass[referee]{aa}
\usepackage{epsfig}
\begin{document}

\newcommand{\gsim}{\hbox{\rlap{$^>$}$_\sim$}}
  \thesaurus{06;  19.63.1}

\authorrunning{S. Dado, A. Dar \& A. De R\'ujula}
\titlerunning{The afterglow of GRB 011121}
\title{Is there a 1998bw-like supernova in the afterglow of
gamma ray burst 011121?}

\author{Shlomo Dado$^{^1}$, Arnon Dar$^{^1}$ and
A. De R\'ujula$^{^2}$}
\institute{1. Physics Department and Space Research Institute, Technion\\   
               Haifa 32000, Israel\\
           2. Theory Division, CERN, CH-1211 Geneva 23, Switzerland}

\maketitle

\begin{abstract}

We use the very simple and successful Cannonball Model (CB) of gamma ray   
bursts (GRBs) and their afterglows (AGs) to analyze the observations of   
the strongly extinct optical AG of the relatively nearby GRB 011121,
which were made with
ground-based telescopes at early times, and with the HST at later time. We   
show that GRB 011121 was indeed associated with a 1998bw-like supernova at   
the GRB's redshift, as we had specifically predicted for this GRB before 
the supernova could be observed.

\end{abstract}

The identity of the progenitors of gamma ray bursts is still unknown,
but it is becoming clearer and clearer.  It
has been suggested that GRBs are produced by highly relativistic jets   
(e.g., Shaviv and Dar 1995; Dar 1998), mostly in
core collapse supernovae like SN1998bw
(Dar and Plaga 1999; Dar 1999a; Dar and De R\'ujula 2000 and references   
therein).  Possible evidence for a SN1998bw-like contribution to a GRB    
afterglow
(Dar 1999a; Castro-Tirado \& Gorosabel 1999) was first found by Bloom et   
al.~(1999a) for GRB 980326, but its unknown redshift prevented a
definite conclusion. The AG of GRB 970228 (located at redshift $\rm
z=0.695$) appears to be overtaken by a light curve akin to that of
SN1998bw (located at $\rm z_{bw}=0.0085$), when properly scaled by their   
differing redshifts (Dar 1999b; Reichart 1999; Galama et al.~2000).
Evidence of similar associations was found for GRB 990712 (Hjorth et
al.~2000; Sahu et al.~2000;  Bjornsson et al.~2001), GRB 980703 (Holland   
et al.~2000), GRB 000418 (Dar and De R\'ujula 2000), GRB 991208
(Castro-Tirado et al.~2001), GRB 970508 (Sokolov et al. 2001), GRB
000911 (Lazzati et al. 2001; Dado et al. 2002a) and GRB 010921
(Dado et al. 2002b).

Unlike supernovae of type Ia (SNe Ia), core-collapse supernovae (SNe   
II/Ib/Ic) are far
from being standard candles. But if their explosions are fairly asymmetric   
---as they would be if a fair fraction of them emit two opposite jets of   
cannonballs---
much of the variability could be a reflection of the varying angles
from which we see their non-spherically expanding shells. Exploiting this   
possibility to its extreme, i.e., using SN1998bw as an ansatz standard   
candle, Dar and De R\'ujula (2000), Dado et al.  (2001a) and De R\'ujula   
(2002) have shown that the optical AG of {\it all} relatively nearby GRBs   
with known redshift (all the ones with $\rm z<1.12$) contain evidence   
or  clear
hints for a SN1998bw-like contribution to their optical AG, suggesting   
that most ---and perhaps all--- of the long duration GRBs are associated   
with 1998bw-like supernovae (in the more distant GRBs,
the ansatz standard candle could not be seen, and
it was not seen).  However, in several of the above cases, lack of
spectral information and multicolour photometry and the uncertain
extinction in the host galaxy prevented a firm conclusion. Thus,
every new instance is still interesting, it will take a few more clear cases 
to reach a generally accepted conclusion.

On Nov. 21, 18:47:21 UT a very bright GRB (011121) was simultaneously   
detected and localized by BeppoSAX (Piro 2001a,b) and IPN (Hurley et al.   
2001).  Its optical afterglow was first detected 10.3 hours after the   
burst in the R-band (Krzysztof et al. 2001).  In further observations   
four additional R-band spectral energy densities were reported
 for the fading source during the first two days after burst (Stanek et 
al. 2001a,b,c), as well as a possible redshift (Infante et al. 2001b) for 
the host galaxy: $\rm z=0.36$. A candidate for this host galaxy was
detected 0.5'' (approximately 2.5 kpc) from the GRB's location (Stanek et 
al. 2001a). As was pointed out by Stanek et al. (2001b), the relatively 
low redshift and the afterglow's fast decay made GRB 011121 an attractive 
search-target for a possible GRB/SN association.

In order to urge and assist the search for a SN in the AG of GRB 011121, 
we used the Cannonball Model of GRBs (Dado et al. 2001a and references 
therein) and the early time R-band observations, recalibrated by the
photometry of Olsen et al. (2001), to predict the late time behaviour of 
the AG in the BVRI bands and to demonstrate that {\it ``The inescapable 
conclusion is that the supernova associated with GRB 011121 will, at about 
day 20 after burst, tower in the BVRI bands above the proper GRB
afterglow''} (Dado et al. 2001b) in spite of the strong extinction in the 
host galaxy and in ours.

Late-time ground-based optical observations were made with the the 6.5m   
Magellan telescope, and an
extensive space-based monitoring campaign was made with the
Hubble Telescope (HST). Indeed, from the first observations on December 4,   
2001, Garnavitch et al. (2002) concluded that the AG of GRB 011121 shows   
the anticipated SN1998bw-like contribution. Price and coauthors concluded   
from observations with HST that {\it ``This curious bump is
inconsistent with
an underlying SN similar to SN 1998bw''} one day (Bloom et al. 2002a), and   
{\it ``It appears that the case for an underlying SN in GRB 011121 is well   
established''} the day after (Kulkarni et al. 2002).  As these authors   
caution,
the conclusions (in the standard fireball paradigm) are affected by possible 
jet breaks (whose position and sharpness cannot be easily predicted).

In this letter we use the Cannonball Model  to estimate the
extinction  in the host galaxy of GRB 011121, and
to predict its late time optical AG\footnote{In the CB model,
like in the data, there are no sharp temporal
breaks; and there is nothing sufficiently
specific and conspicuous to be called a break time nor, therefore,
 any uncertainty associated with it.}.
We show that the evidence is clear: this GRB was indeed
associated with a standard-candle
1998bw-like supernova at $\rm z=0.36$, the GRB's redshift.

\section*{The CB model}

In the CB model, long-duration GRBs and their AGs are produced in core   
collapse supernovae by jets of highly relativistic ``cannonballs'' that   
pierce through the supernova shell. The AG --the persistent radiation in   
the direction of an observed GRB-- has three origins: the ejected CBs, the   
concomitant SN explosion, and the host galaxy. These components are
usually unresolved in the measured ``GRB afterglows'', so that the
corresponding light curves and spectra are the cumulative energy flux   
density:

\begin{equation}
\rm    F_{AG}=F_{CBs}+F_{SN}+F_{HG}\, .
\label{sum}
\end{equation}
The contribution of the candidate host galaxy depends on the
angular aperture of the observations and it is usually determined
by late time observations when the CB and SN contributions become
negligible.

 Let the energy flux
density of SN1998bw at redshift $\rm z_{bw}=0.0085$ (Galama et al. 1998)   
be $\rm F_{bw}[\nu,t]$. For a similar SN placed at a redshift $\rm z$:   
\begin{eqnarray}
{\rm F_{SN}[\nu,t] = } && {\rm{1+z \over 1+z_{bw}}\;
{D_L^2(z_{bw})\over D_L^2(z)}}\, \times\nonumber \\ &&
{\rm
F_{bw}\left[\nu\,{1+z \over 1+z_{bw}},\;t\, {1+z_{bw} \over 1+z}\right]\;   
A(\nu,z)}\, ,
\label{bw}
\end{eqnarray}
 where $\rm D_L(z)$ is the
luminosity distance\footnote{The cosmological parameters we use
are: $\rm H_0=65$ km/(s Mpc), ${\rm \Omega_M}=0.3$ and ${\rm
\Omega_\Lambda}=0.7$.} and $\rm A(\nu,z)$ is the extinction along the line   
of sight.

The contribution of a jet of CBs to the GRB afterglow
at ``late'' times ($\rm t>1$ day) is given by
(Dado et al. 2001a):
\begin{equation}
\rm
F_{CB}=f \; [\gamma(t)]^{3\alpha-1}\;[\delta(t)]^{3+\alpha}\,
\nu^{-\alpha} ,
\label{fluxdensity2}
\end{equation}
where $\rm f$ is a normalization constant (see Dado et al. 2001a
for its theoretical estimate),
$\rm \alpha$ is
the spectral index of the electron synchrotron radiation, $\rm
\gamma(t)$ is the Lorentz factor of
the CB and $\rm \delta(t)$ is its Doppler factor:
\begin{equation}
\rm
\delta\equiv\rm{1\over\gamma\,(1-\beta\cos\theta)}
\simeq\rm {2\,\gamma\over (1+\theta^2\gamma^2)}\; ,
\label{doppler}
\end{equation}
whose approximate expression is valid for small observing angles
$\theta\ll 1$, and $\gamma\gg 1$:
the domain of interest for GRBs.
For an interstellar medium of constant baryon density $\rm n_p$, the   
Lorentz
factor, $\rm\gamma(t)$ is given by (Dado et al. 2001a):
\begin{eqnarray}
\rm \gamma&=&\rm\gamma(\gamma_0,\theta,x_\infty;t)
=\rm {1\over B} \,\left[\theta^2+C\,\theta^4+{1\over C}\right]\nonumber\\   
\rm C&\equiv&\rm
\left[{2\over B^2+2\,\theta^6+B\,\sqrt{B^2+4\,\theta^6}}\right]^{1/3}   
\nonumber\\
\rm B&\equiv&\rm
{1\over \gamma_0^3}+{3\,\theta^2\over\gamma_0}+
{6\,c\, t\over  (1+z)\, x_\infty}
\label{cubic}
\end{eqnarray}
where $\gamma_0=\gamma(0)$, and
\begin{equation}
\rm
x_\infty\equiv{N_{CB}\over\pi\, R_{max}^2\, n_p}
\label{range}
\end{equation}
characterizes the CB's slow-down in terms of
$\rm N_{CB}$: its baryon number, and $\rm R_{max}$:
its radius (it takes a distance $\rm x_\infty/\gamma_0$ for
the CB to half its original Lorentz factor).

The selective extinction, $\rm A(\nu,t)$ in Eq.~(\ref{bw}),
 can be estimated from the difference
between the observed spectral index and the one
expected\footnote{The time dependence of $\alpha$
is analized in detail in Dado et al. (2002c). The CB model predicts,   
and the data
confirm with precision, the predicted gradual evolution of  
$\rm\alpha(t)$ in
the first day or
two to the constant value $\approx 1.1$ observed in all ``late'' AGs
(Dado et al. 2001a).}
in the  CB model ($\alpha\approx 0.5$ at  $\rm t<1$ day,
and $\alpha\approx 1.1$ after a couple of
days). From the early-time relative intensities in the B and V bands of   
the AG of GRB 011121 we obtain a total selective extinction  of
$\rm E(B-V)=0.615\pm 0.07$ magnitudes along the line of sight to  
GRB011121.
Most of this extinction is due to dust in the Galaxy, $\rm E(B-V)=0.50$ 
(Schlegel et al. 1998) in the direction of GRB 011121. The total
selective extinction yields an estimated attenuation factor $\rm
A(\nu,z)=0.18 $ in the V band ($\rm A_V\approx 3.05\, E(B-V)=1.87$
magnitude, Whittet 1991). From the early-time relative intensities in the   
I, R, V and B bands we deduce the attenuation factors
$\rm
A(\nu,z)\sim 0.26\,, 0.22\,,0.18~and~ 0.11$,
to be used in Eq.~(\ref{bw}) to estimate the expected spectral energy   
densities of SN1998bw at the position of GRB 011121.

\begin{figure}[]
\hskip 2truecm
\vspace*{2cm}
\hspace*{-2.6cm}
\epsfig{file=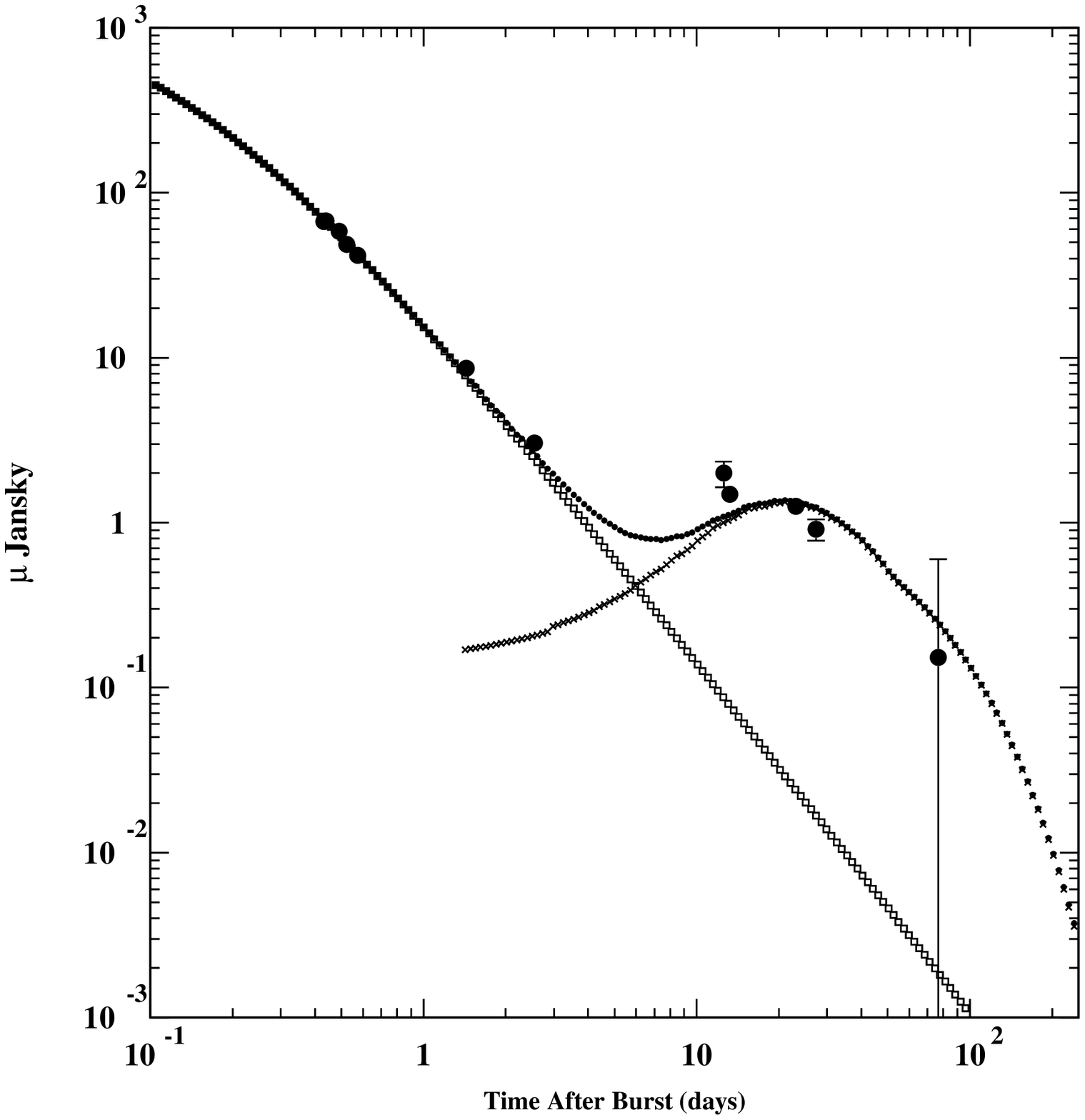, width=15cm}
\caption{Comparisons between our fitted R-band afterglow
and the observations  for GRB 011121 at $\rm z=0.36$.
The CB's AG (the line of squares) is given by Eqs.~(\ref{fluxdensity2}) 
to  (\ref{range}). The observations are not corrected
to eliminate the effect of extinction, so that
the   theoretical contribution  from a 1998bw-like supernova placed at the  
GRB's  redshift,  Eq.~(\ref{bw}), indicated by a line of crosses,
is corrected by the corresponding estimated extinction factor.
The contribution of the host galaxy has been subtracted.}
\label{figr1121}
\end{figure}

\begin{figure}[]
\hskip 2truecm
\vspace*{2cm}
\hspace*{-2.6cm}
\epsfig{file=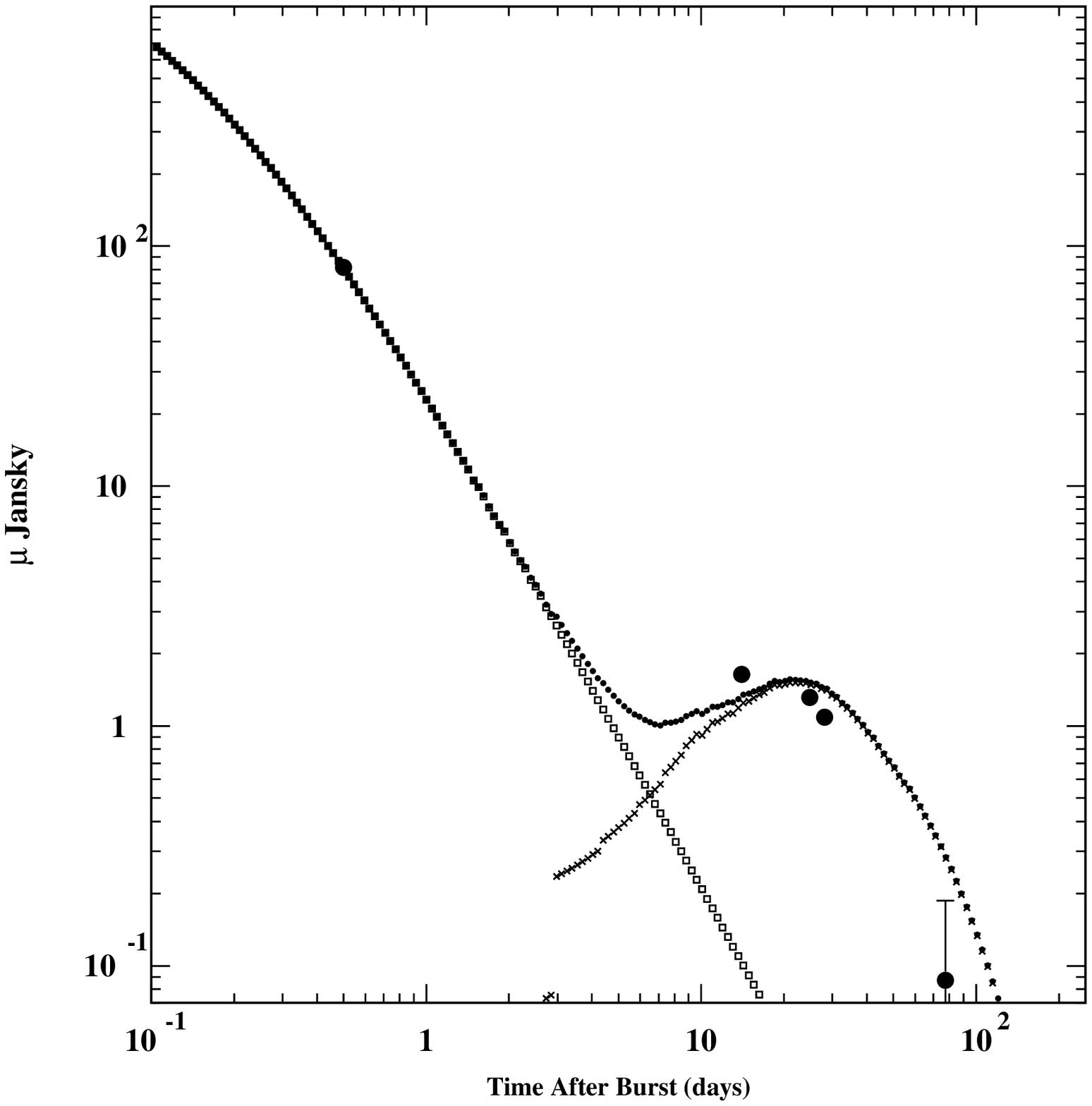, width=15cm}
\caption{Comparisons between our fitted I-band afterglow
and the observations  for GRB 011121 at $\rm z=0.36$.
The CB's AG (the line of squares) is given by Eqs.~(\ref{fluxdensity2}) 
to  (\ref{range}). The observations are not corrected
to eliminate the effect of extinction, so that
the   theoretical contribution  from a 1998bw-like supernova placed at the  
GRB's  redshift,  Eq.~(\ref{bw}), indicated by a line of crosses,
is corrected by the corresponding estimated extinction factor.
The contribution of the host galaxy has been subtracted.}\label{figi1121}   
\end{figure}

\begin{figure}[]
\hskip 2truecm
\vspace*{2cm}
\hspace*{-2.6cm}
\epsfig{file=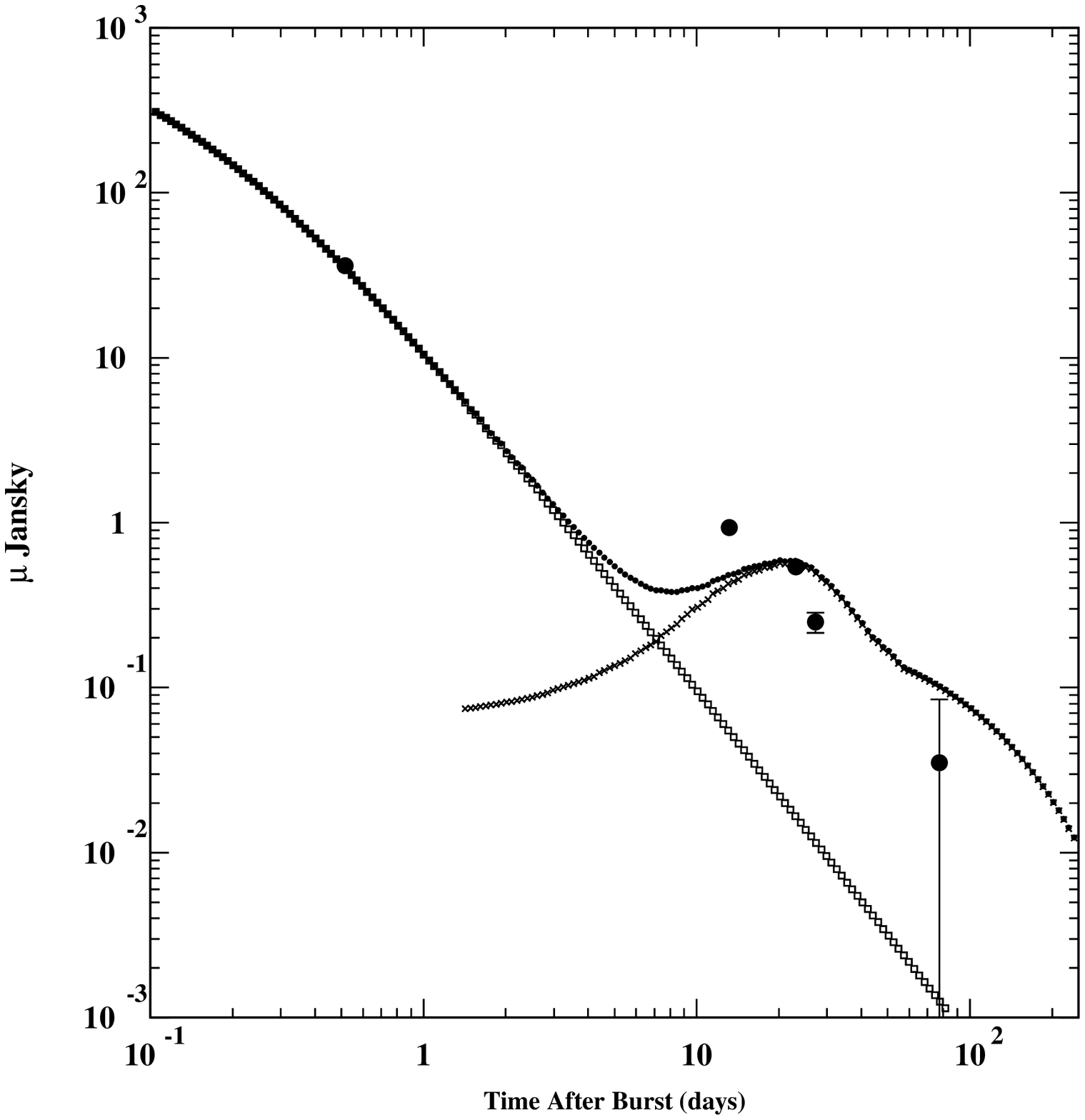, width=15cm}
\caption{Comparisons between our fitted V-band afterglow
and the observations  for GRB 011121 at $\rm z=0.36$.
The CB's AG (the line of squares) is given by Eqs.~(\ref{fluxdensity2}) 
to  (\ref{range}). The observations are not corrected
to eliminate the effect of extinction, so that
the   theoretical contribution  from a 1998bw-like supernova placed at the  
GRB's  redshift,  Eq.~(\ref{bw}), indicated by a line of crosses,
is corrected by the corresponding estimated extinction factor.
The contribution of the host galaxy has been subtracted.}
\label{figv1121}
\end{figure}

\begin{figure}[]
\hskip 2truecm
\vspace*{2cm}
\hspace*{-2.6cm}
\epsfig{file=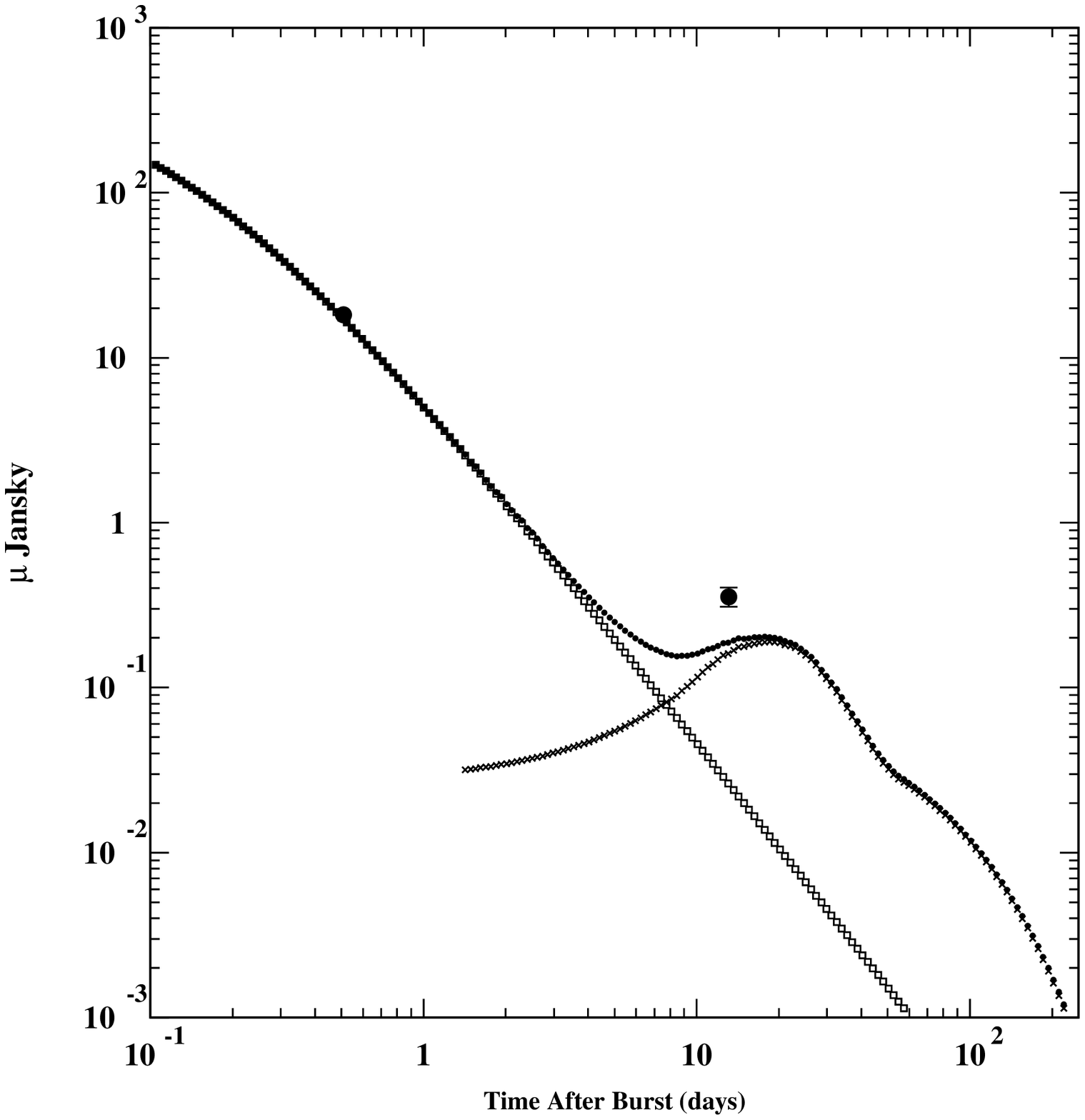, width=15cm}
\caption{Comparisons between our fitted B-band afterglow
and the observations  for GRB 011121 at $\rm z=0.36$.
The CB's AG (the line of squares) is given by Eqs.~(\ref{fluxdensity2}) 
to  (\ref{range}). The observations are not corrected
to eliminate the effect of extinction, so that
the   theoretical contribution  from a 1998bw-like supernova placed at the  
GRB's  redshift,  Eq.~(\ref{bw}), indicated by a line of crosses,
is corrected by the corresponding estimated extinction factor.
The contribution of the host galaxy has been subtracted.}
\label{figb1121}
\end{figure}

We assume, for the late-time AG
of GRB 011121, a spectral slope $\alpha = 1.1$, compatible with that of 
all other GRB AGs (Dado et al. 2001a). The rest of the fitted parameters  
are: $\rm \gamma_0 = 1222$, $\rm \theta = 0.104\, mrad$, and $\rm
x_\infty = 0.83\, Mpc$. The resulting lat-time R-band light curve
is presented in Fig.~(\ref{figr1121}). The contribution of the host  
galaxy
has been subtracted. This contribution
---$\rm F_{HG}\approx 2\, \mu J$ $(\rm R_{host}\approx 23)$
to the early-time measurement,
and
$\rm 0.127\pm 0.026\, \mu Jy$   to the late-time HST measurements---
is a  rough estimate: its
true value depends, respectively, on the angular aperture of the
observations and on the unknown extinction of the host galaxy's
light within this aperture.
In Figs.~(\ref{figi1121}), (\ref{figv1121}) and (\ref{figb1121})
we   present  the CB-model's predictions for the light curves
of the  AG in the IVB bands, with the host galaxy's
contribution subtracted with use of its magnitudes in these bands, as
estimated by Bloom et al. (2002b), which correspond to
$\rm 0.0209\pm 0.059\, \mu Jy$, $\rm 0.087\pm 0.027\, \mu Jy$,
and $\rm 0.098\pm 0.039\, \mu Jy$, respectively.
The theoretical
contribution of an unextinct SN1998bw at redshift 0.36 to the BVRI
bands was reduced by
the attenuation factors $\rm A(\nu,z)=0.11,$ 0.18, 0.22, and 0.26,
due extinction by dust in the host galaxy and in ours, as estimated above.   

\begin{figure}[]
\hskip 2truecm
\vspace*{2cm}
\hspace*{-2.6cm}
\epsfig{file=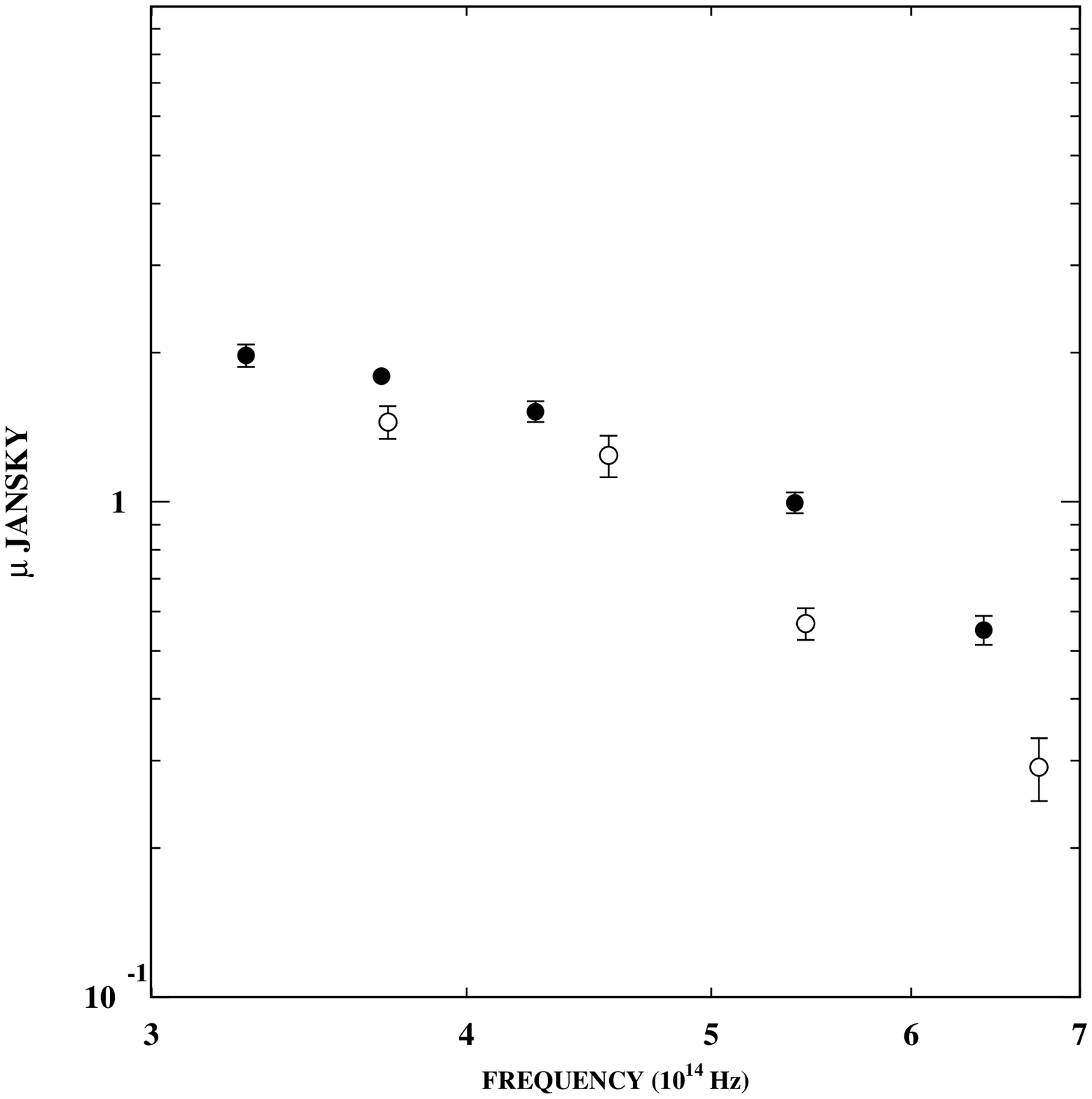, width=15cm}
\caption{The spectral energy density of
the afterglow of GRB  011121 at days 13-14,
at which the SN contribution is fairly dominant.
The filled circles are the HST data of Bloom et al. (2002b),
they are not corrected to eliminate the effect of extinction.
The open circles are the CB-model's predictions: the CB AG
plus a contribution from the host galaxy as estimated by Bloom et al.
(2000b) and a  SN1998bw-like contribution, properly redshifted in its
time- and frequency-dependence, and corrected for extinction,
all as in Eq.~(\ref{bw}).}
\label{fig131121}
\end{figure}

\begin{figure}[]
\hskip 2truecm
\vspace*{2cm}
\hspace*{-2.6cm}
\epsfig{file=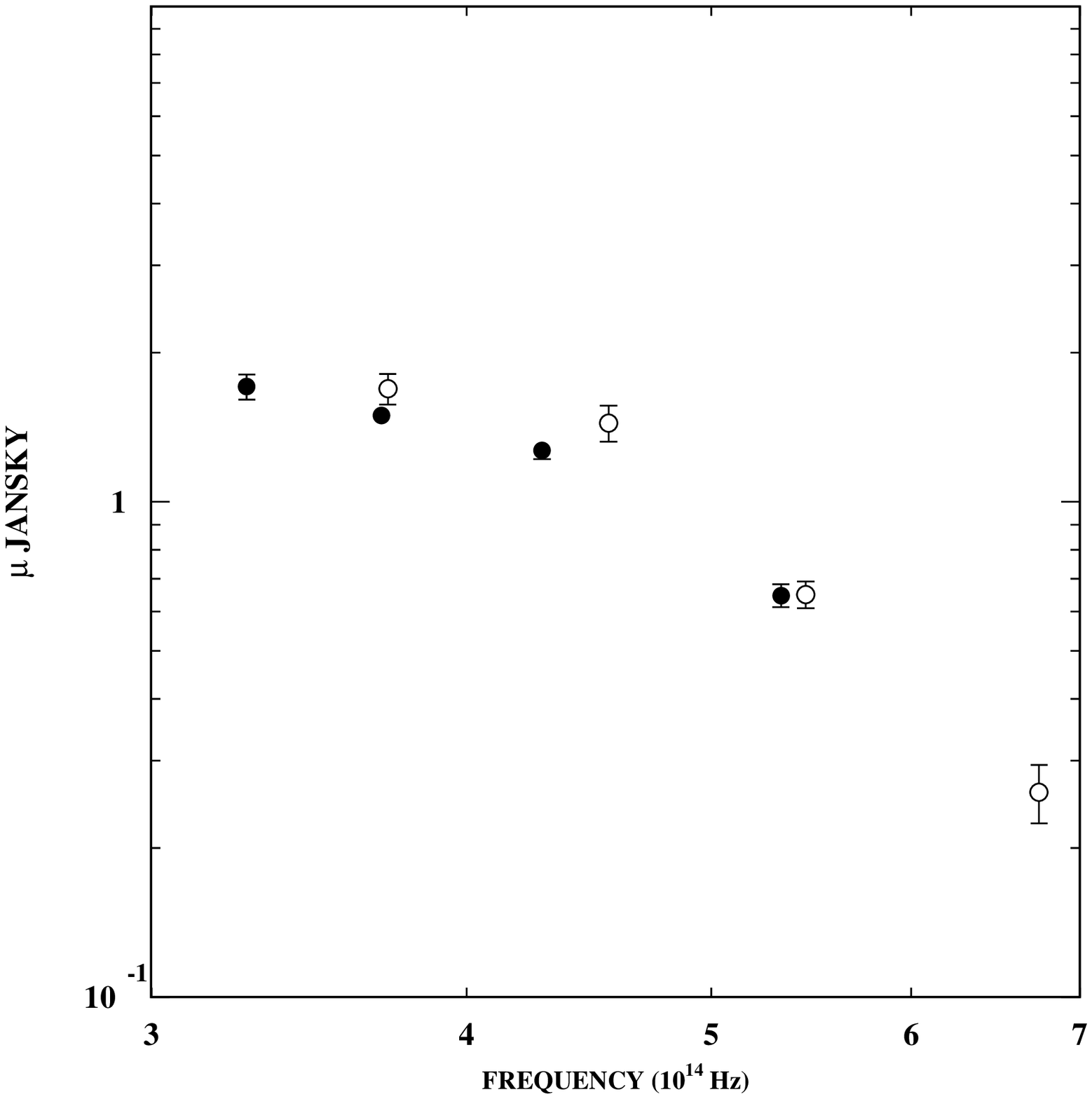, width=15cm}
\caption{The spectral energy density of
the afterglow of GRB  011121 at days 23-24,
at which the SN contribution is fairly dominant.
The filled circles are the HST data of Bloom et al. (2002b),
they are not corrected to eliminate the effect of extinction.
The open circles are the CB-model's predictions: the CB AG
plus a contribution from the host galaxy as estimated by Bloom et al.
(2000b) and a  SN1998bw-like contribution, properly redshifted in its
time- and frequency-dependence, and corrected for extinction,
all as in Eq.~(\ref{bw}).}  \label{fig231121}
\end{figure}

\begin{figure}[]
\hskip 2truecm
\vspace*{2cm}
\hspace*{-2.6cm}
\epsfig{file=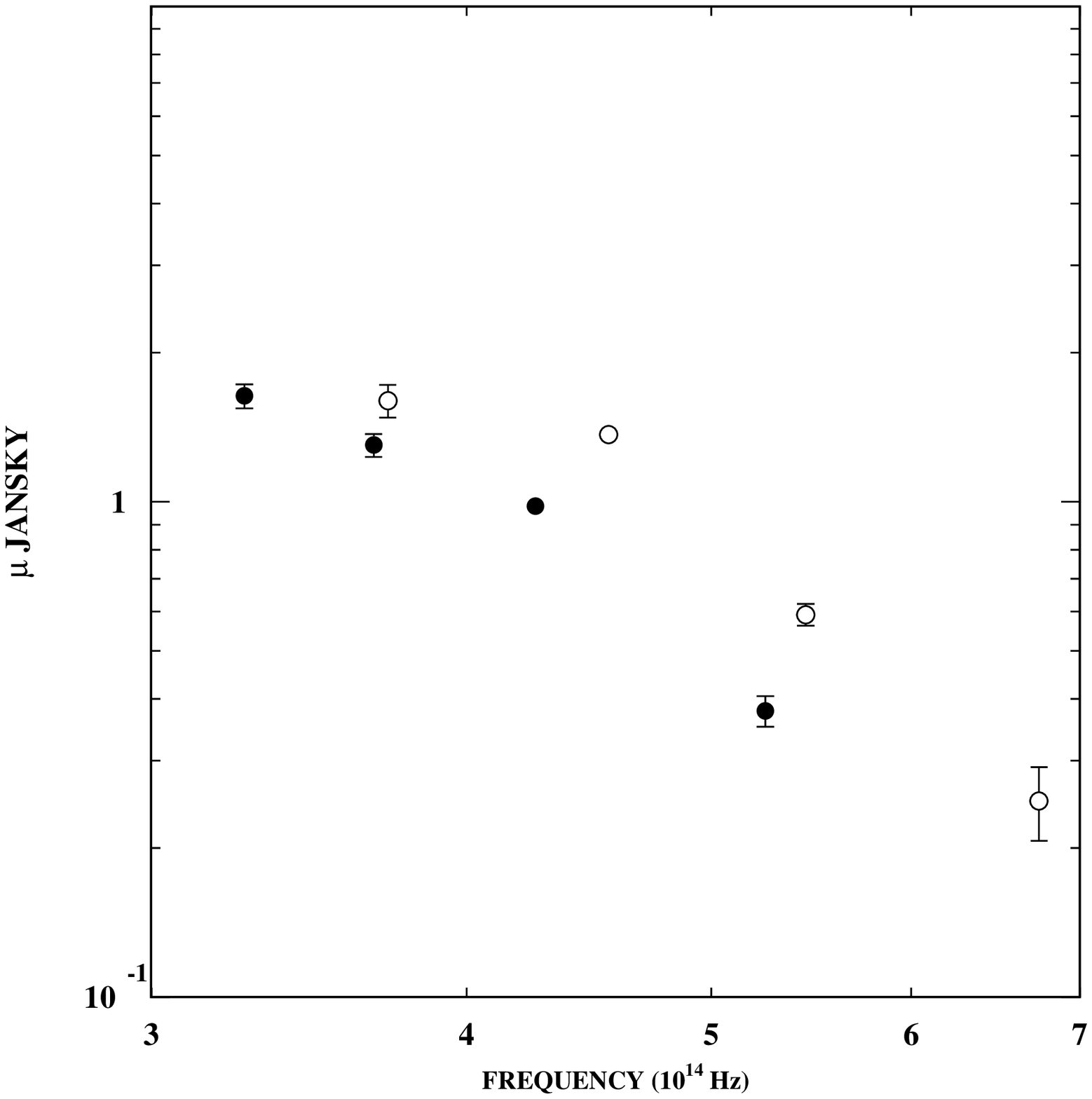, width=15cm}
\caption{The spectral energy density of
the afterglow of GRB  011121 at days 27-28,
at which the SN contribution is fairly dominant.
The filled circles are the HST data of Bloom et al. (2002b),
they are not corrected to eliminate the effect of extinction.
The open circles are the CB-model's predictions: the CB AG
plus a contribution from the host galaxy as estimated by Bloom et al.
(2000b) and a  SN1998bw-like contribution, properly redshifted in its
time- and frequency-dependence, and corrected for extinction,
all as in Eq.~(\ref{bw}).}

\label{fig271121}
\end{figure}

\begin{figure}[]
\hskip 2truecm
\vspace*{2cm}
\hspace*{-2.6cm}
\epsfig{file=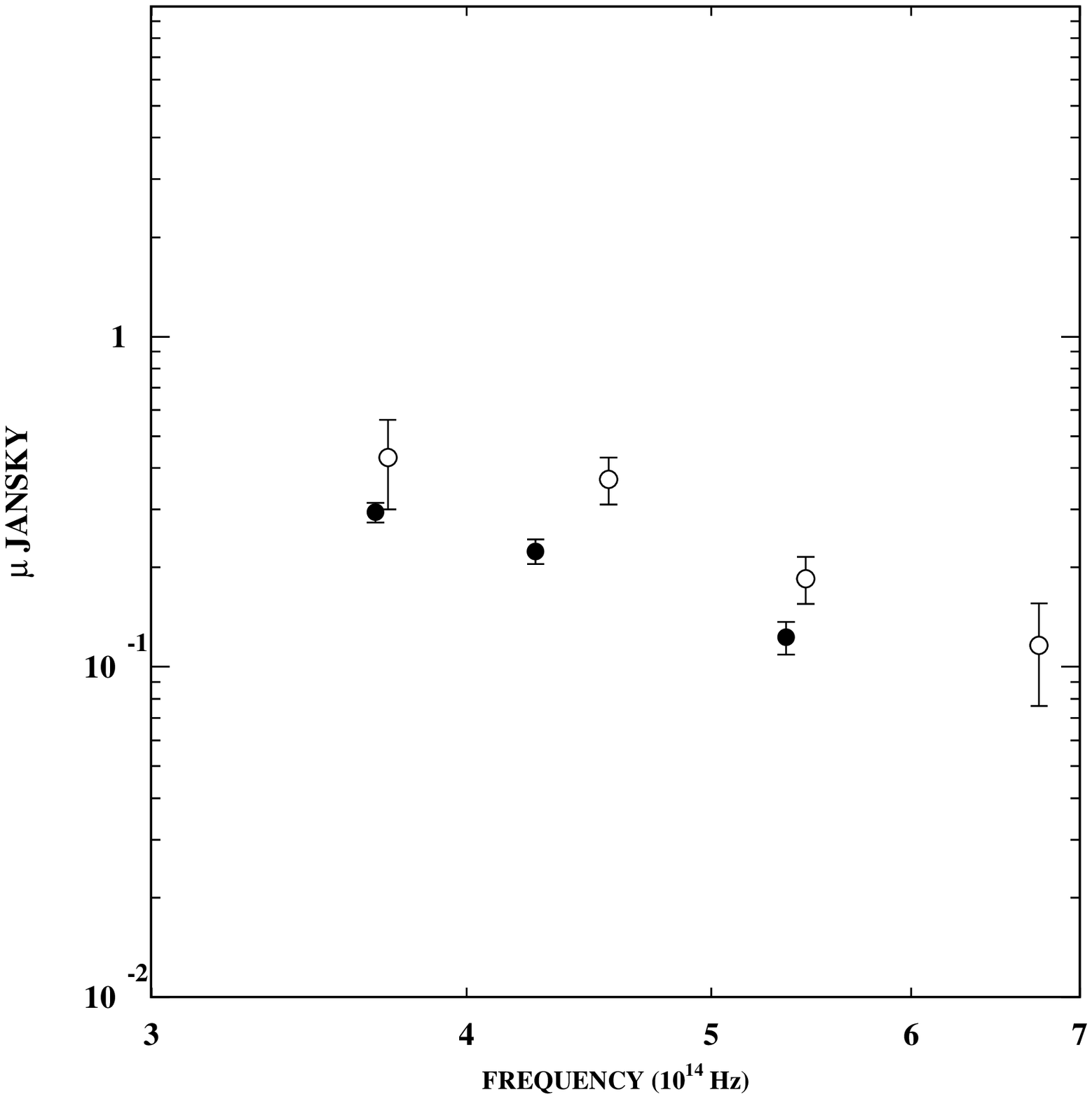, width=15cm}
\caption{The spectral energy density of
the afterglow of GRB  011121 at days 76-77,
at which the SN contribution is fairly dominant.
The filled circles are the HST data of Bloom et al. (2002b),
they are not corrected to eliminate the effect of extinction.
The open circles are the CB-model's predictions: the CB AG
plus a contribution from the host galaxy as estimated by Bloom et al.
(2000b) and a  SN1998bw-like contribution, properly redshifted in its
time- and frequency-dependence, and corrected for extinction,
all as in Eq.~(\ref{bw}).}
\label{fig761121}
\end{figure}

The agreement between theory and observations
in Figs~(\ref{figr1121}) to (\ref{figb1121}) is almost surprisingly good,   
in view of the large observational uncertainties and the theoretical
approximations. A similarly strong correspondence between  predictions 
and  observations is shown
in Figs.~(\ref{fig131121}) to (\ref{fig761121}),
where we compare the predicted late-time spectral energy
distribution, which is to a large extent
dominated by the SN1998bw-like contribution,  with the HST
observations (Bloom et al. 2002b) on days 13-14, 23-24, 27-28 and 76-77.

\section*{Conclusion}

In Dado et al. (2001b) we used the early afterglow data for GRB 011121 
to predict the later AG, in particular the
presence of a SN1998bw-like supernova transported to the GRB's
redshift, which at the earlier time was still unobservable.
All we have done in this letter is to update and refine these expectations 
with use of the improved observational constraints on absorption in the 
host galaxy, and to compare the prognosis with the early
and late AG data. The 1998bw-like contribution is clearly there,
in all of the light-curves at different frequencies, in all of the
spectra at different times.

In Dar and De R\'ujula (2000) we argued that long-duration
GRBs may all be associated with 1998bw-like supernovae,
and that the diversity of core-collapse SNe may to a large
extent be due to a spread of viewing angles, relative to
the CB-emission axis. In Dado et al. (2001a) we showed how
surprisingly successful the ansatz of an associated supernova
identical to 1998bw was, when confronted with the observations
for optical and X-ray AGs. The afterglows of nearby GRBs discovered after 
these quoted works ---that of GRB 000911, discussed in Dado
et al. (2002a), that of GRB 010921, discussed in Dado
et al. (2002b) and that of GRB 011121, discussed here---
strengthen the conclusion: so far, in all AGs in which a
SN like 1998bw was visible (in practice, in all cases
with redshift $\rm z<1.12$), it was seen. And it was
compatible in magnitude and colour with an  SN1998bw
standard-candle! It goes without saying
that there are no standard candles. It is just that the current data are not 
precise enough to detect significant variations in this particular one. 
But the important fact is that the supernovae allegedly associated with  
all long-duration GRBs (Dado et al. 2001a, and references therein)
{\bf are indeed there}.

\end{document}